\documentclass[aps,prl,12pt,showpacs,superscriptaddress]{revtex4}

\usepackage{graphicx}
\usepackage{amsmath,units}
\usepackage{mathrsfs}
\usepackage{color}

\usepackage{epstopdf}

\definecolor{cream}{RGB}{222,217,201}

\begin{document}

\title{Kinetics of actin networks formation measured by time resolved particle-tracking microrheology.} 

\author{Maayan Levin}
\affiliation{Raymond \& Beverly Sackler School of Chemistry, Tel Aviv University, Tel Aviv 6997801, Israel.}

\author{Raya Sorkin}
\affiliation{Raymond \& Beverly Sackler School of Chemistry, Tel Aviv University, Tel Aviv 6997801, Israel.}

\author{David Pine}
\affiliation{Department of Physics, New York University, NY 10003, USA.}
\affiliation{Department of Chemical \& Biomolecular Engineering, New York University, Brooklyn, NY 11201, USA.}

\author{Rony Granek}
\affiliation{Department of Biotechnology Engineering and NIBN, Ben-Gurion University, Beer Sheva 84105, Israel.}

\author{Anne Bernheim-Groswasser}
\affiliation{epartment of Chemical Engineering and Ilse Katz Institute for Nanoscale Science and Technology Ben Gurion University of the Negev Beer-Sheva 84105, Israel.}

\author{Yael Roichman}
\email{roichman@tauex.tau.ac.il}

\affiliation{School of Chemistry, Tel Aviv
	University, Tel Aviv 6997801, Israel}
\affiliation{School of Physics \& Astronomy, Tel Aviv
	University, Tel Aviv 6997801, Israel}

\date{\today}

\begin{abstract}
	Actin is one of the most studied cytoskeleton proteins showing a very rich span of  structures. It can self-assemble actively into dynamical structures that govern the mechanical properties of the cell, its motility and its division. However, only very few studies characterize the kinetics of the active actin self-assembly process beyond the formation of an entangled network. Here, we follow actin polymerization kinetics and organization into entangled networks using time resolved passive microrheology. We establish a relationship between the initial concentration of monomers, the active polymerization and network formation kinetics, and the viscoelastic properties from the onset of actin polymerization upto the formation of a steady state entangled network. Surprisingly, we find that at high enough initial monomer concentrations the elastic modulus of the forming actin networks overshoots and then relaxes with a -2/5 power law, that we attribute to rearrangements of the network into a steady state structure.  
\end{abstract}

\maketitle

	\section{Introduction}
	The mechanisms governing self-organization processes have long been a subject of interest for a large community of physicists, chemists and engineers.
	Of special interest are biological systems, as the cell cytoskeleton, that self-assemble and organize while consuming chemical energy. These systems self-organize continuously with little errors on many length scales, and  in highly noisy conditions. The most noted example being the coding and transcription of genetic information \cite{Gout2013}. The actin cytoskeleton plays a major role in maintaining the cell shape and mechanical properties  \cite{Cooper1991,Pollard1208,Blanchoin2014}. Actin filaments are also the substrate for the action of the myosin motor protein and together they take a prominent role in many active cellular processes such as cell motility and cell division \cite{gardel2013,rev2017}. In order to characterize the kinetics of the self-assembly of actin monomers into actin filaments and their organization into actin networks, it is necessary to observe both the nucleation and growth of actin polymer chains as well as the network formation process. Previous experiments investigated actin polymerization kinetics by using in-vitro model systems, in order to isolate the relevant variables of the network formation process \cite{Anne2006,selfass1,selfass2}.
	So far, little is known about the dynamics of this process due to the sub-diffraction-limit size of the actin filaments at the initial stages, and the difficulty to visually examine this process in real time at intermediate and final stages in bulk solution.
	
	The kinetics of actin networks formation was mostly investigated by pyrene assay \cite{Cooper1983,peryne1983,poly2013}, which is based on the enhanced fluorescence of pyrene-conjugated actin that occurs during polymerization. When pyrene with monomeric G-actin forms pyrene F-actin, the intensity of fluorescence increases until it  saturates to a final value when the polymerization process ends. Therefore, this method alone does not give information about the structural and mechanical properties of the forming network, and does not insure that a network has been formed.
	In order to overcome the sub-diffraction limit, super resolution imaging of actin was preformed \cite{super1,Super2,Super3} on dilute suspensions, low density networks and in the cortex of cells \cite{DiegoPRX,KUHN2005}. Super resolution methods are direct and reliable but cannot resolve the evolution of actin networks, which dynamically changes their properties in time scales of milliseconds.  
	
	Here, we characterize the kinetics of self-organization of actin networks at time scales unattainable previously. We do so by starting at different monomeric actin concentrations, using time resolved microrheology. Using tracer particles that are larger than the final mesh size of the formed networks, allows us to follow through the formation and relaxation of the self-organized network until it reaches a steady state. We note that microrheology of actin networks with and without myosin motors was studied extensively \cite{Gijsje2012,Schmidt2017,gardel2011,Lee2009,Martin2015}. However, it was not used previously to obtain the details of polymerization kinetics and network rearrangement.

	\section{Materials and Methods}
	\subsection{Materials}
	G-actin was purified from rabbit skeletal muscle acetone
	powder \cite{1971actin}, with a gel filtration step, stored on
	ice in G-buffer (5 mM Tris HCl, 0.1 mM $\text{CaCl}_2$, 0.2
	mM ATP, 1 mM DTT, 0.01\% $\text{NaN}_3$, pH 7.8) and used
	within two weeks. G-actin concentration was determined by absorbance, using a UV/Visible spectrophotometer (Ultraspec 2100 pro, Pharmacia) in a cuvette with a 1 cm path length and extinction coefficient of $\epsilon_{290} = 26, 460 M^{-1}cm^{-1}$. Polystyrene colloids with
	diameter of $1.5 \mu m$ (Polysciences, Catalog No. 09719-10) were incubated for 2hr before the experiment began with a 10 mg/ml bovine serum albumin (BSA) solution (Sigma) to
	prevent nonspecific binding of protein to the bead surface \cite{valentine2004}.
	Actin polymerization was initiated by adding G-actin at various concentrations to F-buffer solution (5 mM Tris HCl, 1 mM $\text{MgCl}_2$, 0.05 M KCl, 200 $\mu$M EGTA, 1 mM ATP) and gently mixing. The colloidal particles were added before mixing. The initial actin monomer concentration varied between $C_A = 2 - 24 \mu M$, corresponding to a mesh size range of $\xi = 0.95 - 0.25 \mu m$, respectively \cite{SCHMIDT1989}.

	\subsection{Sample preparation}
	Samples were prepared on glass coverslips (40mm in diameter) coated with methoxy-terminated PEG
	(Polyethylene glycol, Mw=5000 g/mol, Nanocs) for 2hr before the experiment began to prevent
	F-actin filaments from sticking to the glass surface. The sample chamber was $\sim150 \mu m$ high and sealed with paraffin wax. Actin polymerization was carried out by mixing the monomers, buffer, and the fluorescent tracer particles together. The sample holder was initially positioned on the microscope. Immediately after mixing, the suspensions were placed on the sample holder and the sample was sealed. This ensured minimal drifts in the sample and enabled us to see the polymerization process from its onset.  The lag time between mixing and recording was measured by a stop watch. In order to avoid wall effects, imaging was done in the solution bulk approximately $60 \mu m$ away from the two glass coverslips. The resulting F-actin networks are well described as networks of semi-flexible polymers, and their mesh size, $\xi=0.3/\sqrt{c_a}$ \cite{SCHMIDT1989}, was easily controlled by controlling the initial monomer concentration $c_a$ ($c_a$ in mg/ml and $\xi$ in $\mu m$). 
	
	\subsection{Fluorescence microscopy and imaging}
	Imaging the motion of the tracer particles within the suspensions was done using an Olympus IX71 epi-fluorescence microscope, at $\lambda = 480$nm with a 40x air objective. We recorded the motion of approximately 250 particles in the field of view using a CMOS video camera (Grasshoper 3, Point Gray) at a frame rate of 30Hz with an exposure time of 10ms. We used data from at least $10^5$ frames per experiment and repeated each experiment three times. Particle tracking was done by conventional video microscopy, using the protocol of Crocker and Grier \cite{Crocker1996} implemented in MATLAB software. 
	
	\subsection{Time resolved microrheology}
	
	The transient mechanical properties of the polymerizing actin solution were measured by one point microrheology (1P microrheology)\cite{Mason1995,Squires2010,Crocker2007} using the following protocol. 
	First we recorded movies of the mixed actin solution with the tracer particles as the actin polymerized. We then extracted the trajectories of the tracer particles throughout the movies. The movies were divided into sections of 30s. For each such a section we extracted the viscoelastic properties of the suspension, assuming it represents the average mechanical properties of the suspension in that time window. Namely, for all trajectories belonging to a given section we calculated the mean squared displacement (MSD) as a function of lag time, according to: $\langle \Delta \vec{r}^{2}(\tau) \rangle = \langle|\vec{r}(t+\tau)-\vec{r}(t)|^2 \rangle_t$, where $\vec{r}$ is the particle position, $\tau$ is the lag time and $t$ is the time since the beginning of the experiment. The averages were done both on $t$ and on the ensemble of tracer particles per section. 
	The shear modulus at time $t$ of the evolving network was calculated using the generalized, frequency-dependent form of the Stokes-Einstein relation. Specifically, the macroscopic Laplace-transformed complex shear modulus $G(s)$ and the Laplace transform of the MSD $ \langle \Delta \tilde{r}^2(s) \rangle $ are related through \cite{Mason1995,Crocker2000,Mason2000}:
	\begin{equation}
	\tilde{G}(s) = \frac{k_BT}{\pi as\langle \Delta \tilde{r}^2(s) \rangle},
	\label{Eq:GSER}
	\end{equation}
    The storage and loss shear modulus are defined as the real and imaginary part of the shear modulus $\tilde{G}(\omega)=G'(\omega)+iG''(\omega)$ and can be obtained  directly from $G(s)$ \cite{Mason1995}. Errors where estimated by the standard error of the mean.
	
	\section{Results}
	\begin{figure}[h!]
		\centering
		\includegraphics[scale=0.25]{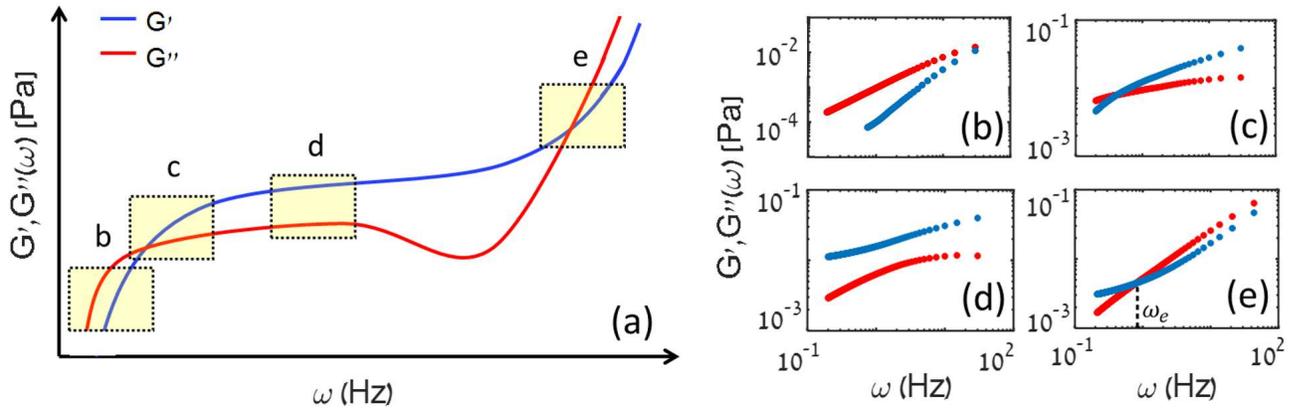}
		\caption{Frequency dependence of the storage and loss modulus. (a) Illustration of the full range of behavior of an entangled semiflexible polymer system. (b-e) measurements of the shear moduli at the same frequency window. We note that in different conditions, i.e. different times and actin monomer concentrations, different regions of (a) are accessible. Experimental conditions: (b) $3\mu M$ actin after 2.5 min of polymerization, (c) $17\mu M$ after 2.5 min of polymerization, (d) $17\mu M$ actin after 3.5min of polymerization and (e) $3\mu M$ after 60min of polymerization respectively (b-e). Although the frequency is identical for all measurements, they exhibit different regions of frequency in illustration (a) due to the reptation time increase (see supplementary Table S2).}
		\label{fig:All_freq_range}
	\end{figure}
\begin{figure}[h!]
		\centering
		\includegraphics[scale=0.2]{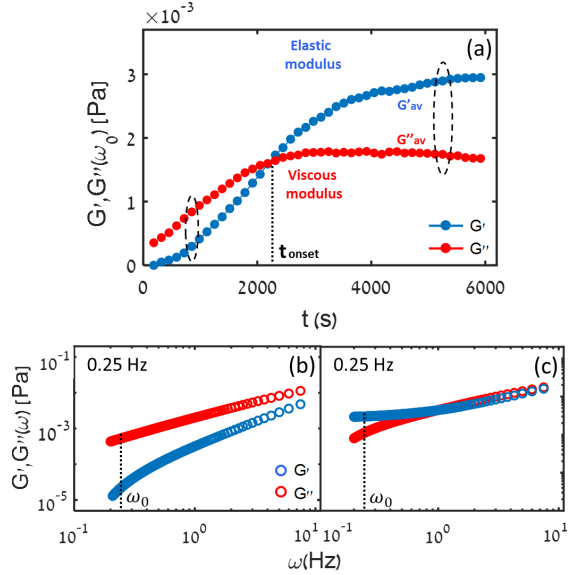}
		\caption{A typical plot of the process of $3\mu M$ actin network polymerization as observed from the change of the elastic and loss shear modulus at $\omega_0$ for entanglement time determination (a), and the corresponding $G(\omega)$ graphs at the beginning (b) and at the end (c) of network's formation.}
		\label{fig:Gcalc}
	\end{figure}

	In order to capture in detail the whole process of self-assembly of the actin filaments and the self-organization of the actin networks, we tuned the initial concentration of actin monomers according to previous reports of pyrene based experiments \cite{peryne1983}. Our primary concern was to ensure that the polymerization and network formation process occur in timescales that are amenable for time resolved microrheology, i.e., in the range of several to tens of minutes. For each monomer concentration, we record the movement of the tracer particles during the polymerization process. We then segmented the movie into sections of 30s and calculate the shear modulus  as a function of frequency (see Methods for details). 
	Since we work with a given time window, we can access a given range of frequencies, $0.2 Hz<\omega<30Hz$. These frequencies reflect different sections of the functional dependence of $G'$ and $G''$ on frequency, depending on actin concentration and/or polymerization time (e.g., the time passed since the initiation of the polymerization). Fig.~\ref{fig:All_freq_range} illustrates $G'(\omega)$ and $G''(\omega)$ at a full frequency range according to the tube theory of entangled polymer dynamics  \cite{Doi_Edwards1,Doi_Edwards2,Doi_Edwards3,Doi_Edwards4} with corresponding examples from our measurements. The transition between windows (b-e) in Fig.~\ref{fig:All_freq_range} is attributed to the reptation time increase (\dag{ESI}, Table S2) \cite{granek1997}. A full time progression for the same actin monomer concentrations that are plotted in Fig.~\ref{fig:All_freq_range}(b-e), is shown in the \dag{ESI}, Fig. S4. The storage and loss moduli in Fig.~\ref{fig:All_freq_range}(e) exhibit a scaling of 3/4 corresponds to theory \cite{GittesPRL1997}.
	We note here, that the calculated complex shear modulus is only an estimation of the actual shear modulus, since we are far from thermal equilibrium and since the tracer particle is not big enough compared to the  mesh size to provide the true bulk response throughout the experiment\cite{haim}. None-the-less, for the purpose of following the kinetics of the process, it is the change in the shear modulus that is important and not its absolute value. 
	We characterize the temporal evolution of self-organizing actin networks by recording the values of $G'$ and $G"$ at $\omega_0 = 0.25 Hz$ Hz (or $\tau = 4s$) for each time segment since the initiation of polymerization. This is repeated for different values of the initial concentration of actin.
	Ideally, we would characterize the viscoelastic properties of the networks in a well defined region such as the elastic plateau, namely at $\omega->0$. However, this region is not attainable in our experiments. We therefore compromise on a value that allows collection of a sufficient amount of data while at the lowest possible frequency. The values of the shear modulus obtained thus for each segment of the movie are then plotted as a function of time since polymerization initiated (Fig.~\ref{fig:Gcalc}). For all actin concentrations, we observe that at the beginning of polymerization,  $G'$ is lower than $G''$ for all measurable frequencies $\omega$ (including at $\omega_0$), indicating that the actin suspension behaves as a viscous fluid. Later, filaments start to elongate and overlap and we observe a viscoelastic behavior at the lower frequencies. After entanglement of filaments has occurred, $G'$ is higher than $G''$ at low frequencies and lower at high frequencies (as seen in Fig.~\ref{fig:Gcalc}(b,c)). At high actin concentrations, we reach a range in which the actin network is fully formed indicated by the fact that  $G'$ is higher than $G''$ at all frequencies
	
\begin{figure*}[h!]
		\centering
		\includegraphics[scale=0.22]{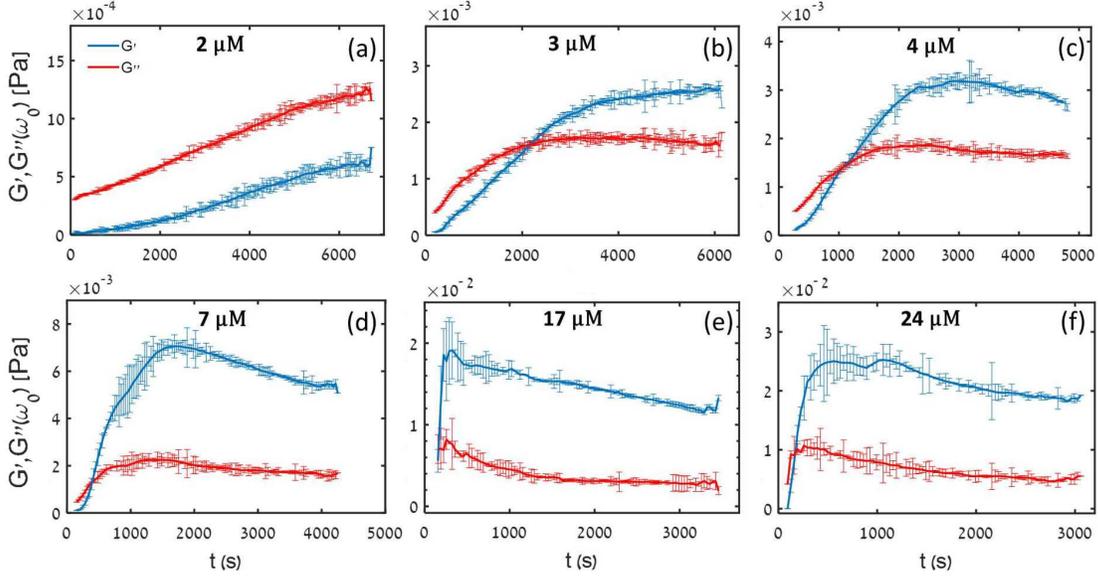}
		\caption{The temporal evolution of $G'$ and $G''$ at $\omega_0$ for actin monomer concentrations varying from 2 to 24 $\mu M$. The shear moduli of $C_A$ = $7-24 \mu M$ exhibit overshoot after network entanglement. A short lag phase observed at the beginning of polymerization for $C_A$ = $2-7 \mu M$.}
		\label{fig:G_CuM}
	\end{figure*}

	The kinetics of the polymerization and self-organization of the actin network can be extracted from the functional dependence of the shear moduli on the time since polymerization initiated Fig.~\ref{fig:Gcalc}(a), where we define the time for the onset of entanglements $t_{\text{onset}}$ to be the time in which $G'(\omega_0)$ becomes larger than $G''(\omega_0)$ (see dashed line in Fig.~\ref{fig:Gcalc}(a)). We also define the averaged shear moduli of the final network as its average value at long times {(Fig.~\ref{fig:Gcalc}(a))}.  The total duration of the polymerization and organization of the actin network is the time in which $G'$ and $G''$ reach their steady state value. Using a range of initial actin monomer concentrations, we are able to study the kinetics of self-organization at different conditions and for networks with different mesh sizes, as shown in Fig.~\ref{fig:G_CuM}. 
	
	As seen in Fig.~\ref{fig:G_CuM}, at very short times and mostly at low concentrations of actin we observe a time delay before the suspension starts to increase it's viscosity. We attribute this lag to the lower probability of nucleation. An abrupt increase in the shear modulus is observed after a few minutes, depending on actin concentration. In this stage the suspension still behaves as a viscous fluid with increasing viscosity, probably reflecting the growth of actin filaments. When the length of the filaments reaches a critical value, the system becomes an entangled network and proceeds to stabilize until reaching a steady state.
	In actin concentrations which are greater than $4\mu M$, a strong overshoot of the networks' storage modulus occurs at the later stages of filament elongation and then relaxes to the steady state value. 
	
\begin{figure}[h!]
		\centering
		\includegraphics[scale=0.22]{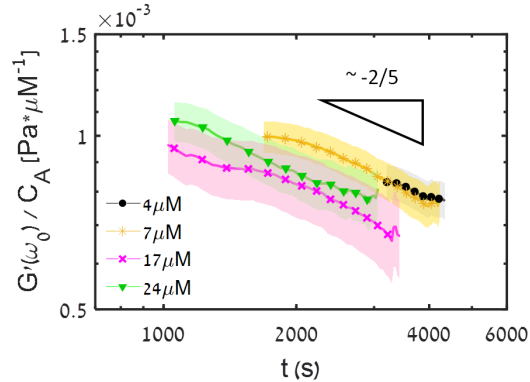}
		\caption{$G'$ Relaxation regions fall on a master curve once $G'$ is normalized by actin monomer concentration and presented as a function of time.}
		\label{fig:GpGpp_master_curve}
	\end{figure}

	Interestingly, the relaxation curves of all samples which exhibited an overshoot collapse into a master curve when normalized by the initial actin monomer concentration \cite{Koenderink2014} with a relaxation power law of -2/5, as shown  in Fig.~\ref{fig:GpGpp_master_curve}. We assume that this effect is related to rearrangements of the filaments within the network, since we eliminated the possibility that it is caused by a lack of ATP (see \dag{ESI} Fig. S1). All curves were taken from G' values that exhibit a frequency dependency corresponds to Fig.~\ref{fig:All_freq_range}(d,e).     
	
	\begin{figure}[h!]
		\centering
		\includegraphics[scale=0.18]{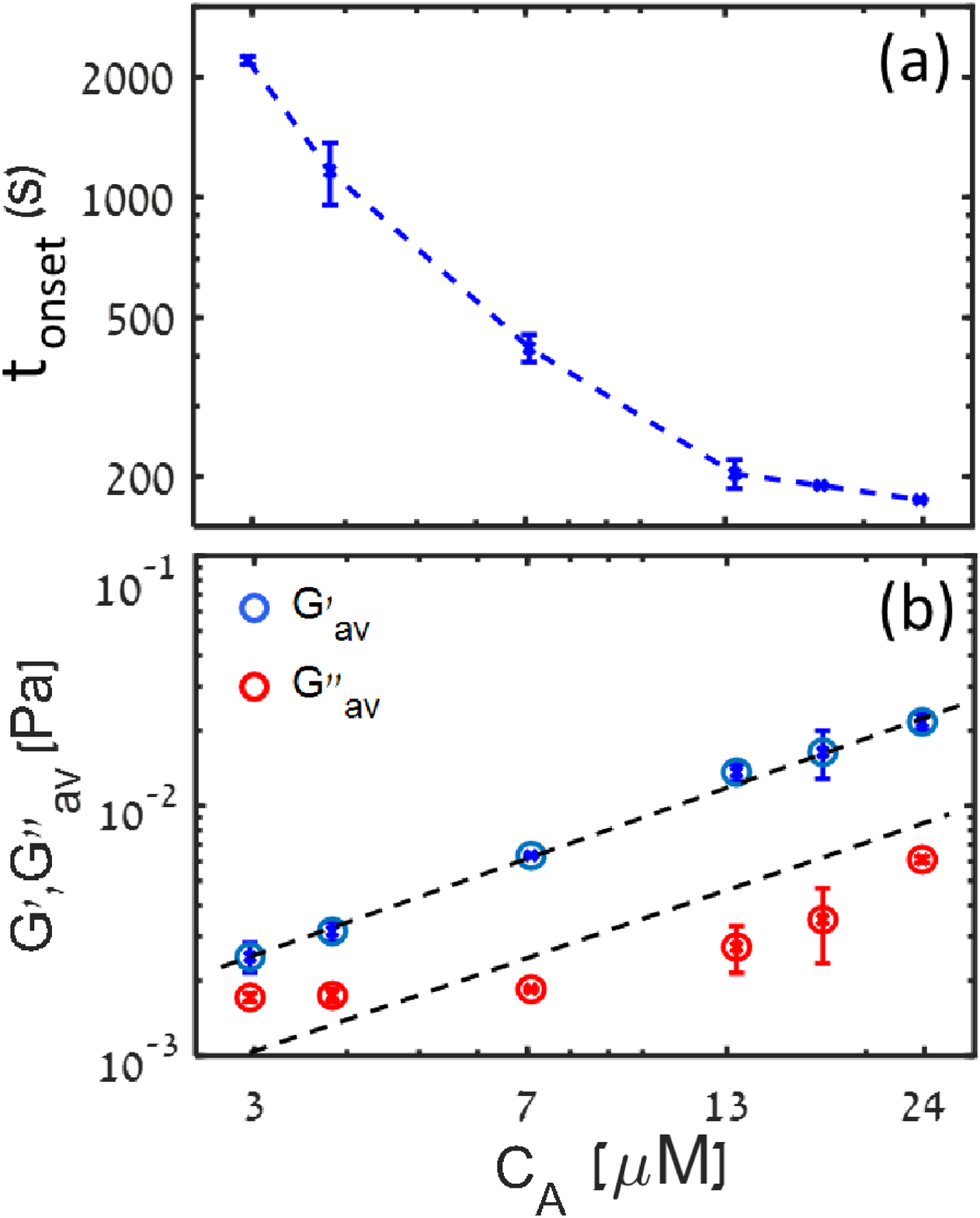}
		\caption{Effect of actin concentration on the entanglement time (a), and the averaged stiffness of the network at long experiment times (b). The dashed lines represent a fit to a linear dependence between the shear moduli and $C_A$}
		\label{fig:t_Gav_CuM}
	\end{figure}
	
	For each curve in Fig.~\ref{fig:G_CuM}, the entanglement time and the averaged shear modulus of the final network were extracted (see Fig.~\ref{fig:t_Gav_CuM}). We find that the lower the monomer concentration is, the longer it takes the solution to form a network. In addition, the final stiffness of the networks increases with actin concentration (Fig.~\ref{fig:t_Gav_CuM}(b)), as expected (consistent with $G'\sim c_A$)\cite{Koenderink2014}.
	
	In order to estimate the filament length during polymerization, we refer to the theory of diffusion of dilute suspension of rigid rods \cite{Briels2003}. This is justified since the actin filaments have a long persistence length of $l_p \sim 17 \mu m$ \cite{Libchaber1993,Howard1993}. The averaged filament length $L$ is thus given by \cite{Briels2003}: 
	\begin{equation}
	L \sim D \sqrt{\frac{\eta_{\text{eff}}/\eta_0-1}{A}},
	\label{Eq:L}
	\end{equation}
	where $A=\frac{4}{3}\frac{1}{ln(1/\phi)}(1-\frac{ln(ln(1/\phi)}{ln(1/\phi)}+\frac{0.6634}{ln(1/\phi)})$, D is the filament width, $\eta_{\text{eff}}$ and $\eta_0$ are the viscosity of the solution and the solvent respectively and $\phi$ is the volume fraction of polymerized actin monomers. It is important to note that this calculation is valid before entanglement occurs only (i.e. until $t_{\text{onset}})$. After the onset of entanglements, the suspension can not be treated as a viscous fluid.   
	
	\begin{figure*}[h!]
		\centering
		\includegraphics[scale=0.3]{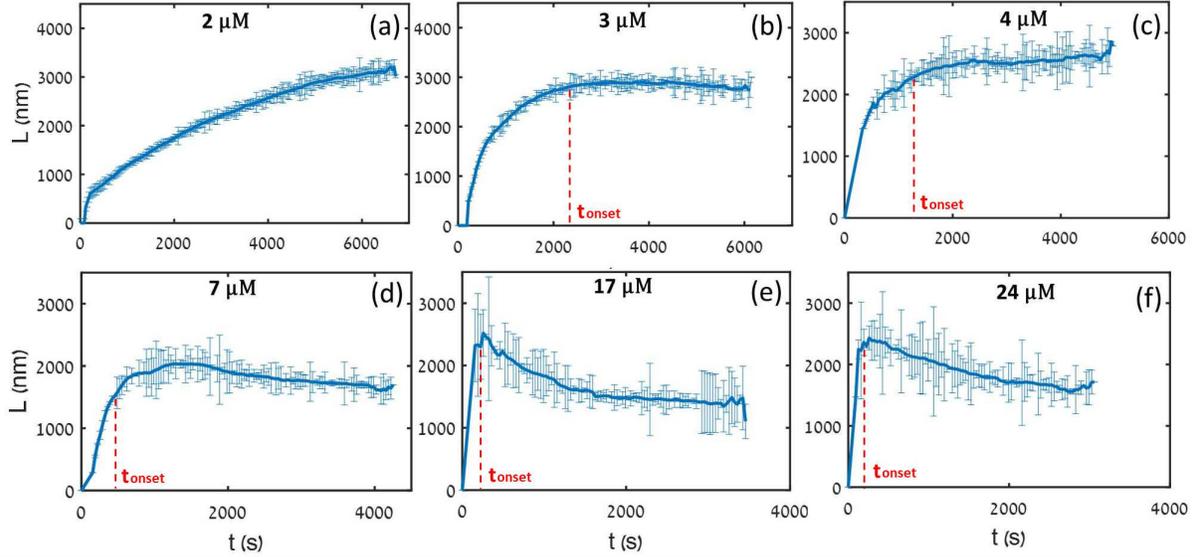}
		\caption{Average filament length (L) as a function of time extracted from Eq.~\ref{Eq:L} for monomer concentrations varying from 2 to 24 $\mu M$. $t_{\text{onset}}$ denotes the time in which entanglement starts, e.g. when the dilute diffusing rod suspension limit is no longer valid. The final average length of filaments decreases with an increase in actin monomer concentration}.
		\label{fig:L_all_err}
	\end{figure*}
	
	We find that the lower the monomer concentration is, the longer the forming filaments are. When the concentration is low, fewer nucleation sites are being formed and the filaments can use the free monomers to elongate more. This is due to the fact that the rate determining step for actin polymerization at these concentration is the nucleation. When the initial concentration of actin monomers increases, there are more nucleation sites, which reduces the amount of available monomers per growing filament. It is important to note that the calculated lengths reflect the relative length in the different networks and not the actual values. This is because they were extracted from a shear modulus obtained by 1P microrheology (which does not reflect the bulk's viscosity). In fact, we expect the actual shear modulus to be higher than measured here, indicating that the actual filament length is longer (see comparison to 2P microrheology in supporting information (Fig. S2) and previous reports\cite{Gardel2003,ent2,Adar2014}). This is especially true at low actin concentrations where the ratio between the mesh size and the diameter of the tracer particles is small \cite{haim}.

\begin{figure}[h!]
		\centering
		\includegraphics[scale=0.16]{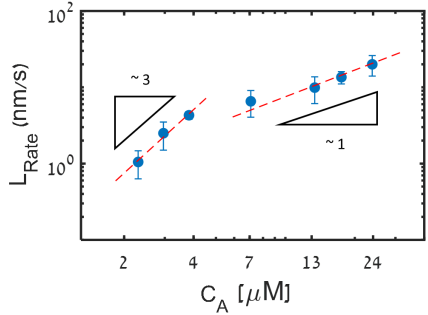}
		\caption{The initial elongation rate $L_{\text{rate}}$ extracted from Fig.~\ref{fig:L_all_err} as a function of actin monomer concentration with fitted power laws of $L_{\text{rate}} \propto [actin]^3$ and $L_{\text{rate}} \propto [actin]$ respectively. }
		\label{fig:rate_c}
	\end{figure}

	The initial elongation rate, $L_{\text{rate}}$ was calculated for each actin concentration at the beginning of the polymerization process in order to derive estimation for the rate law. The dependence of $L_{\text{rate}}$ on the initial actin monomer concentration is shown in Fig.~\ref{fig:rate_c}.
	We observe two distinct regimes, for low actin concentrations $L_{\text{rate}} \propto [actin]^3$ and for higher concentrations $L_{\text{rate}} \propto [actin]$. This is consistent with a nucleation limited growth at low concentrations and diffusion limited growth at higher concentrations. The $L_{\text{rate}} \propto [actin]^3$ power law obtained here might indicate that the nucleus is built of three monomers, which is consistent with previous results that estimate the size of the nucleus to be two to four monomer units \cite{peryne1983}. For high actin concentrations, the reaction rate grows linearly with the concentration as expected, since the elongation rate depends on the collisions rate  \cite{POLLARD2003,Hikoichi1983,KUHN2005}. Consistent results were obtained from the 2P microrheology analysis (see \dag{ESI} Fig. S3). 
	
\begin{figure}[h!]
		\centering
		\includegraphics[scale=0.225]{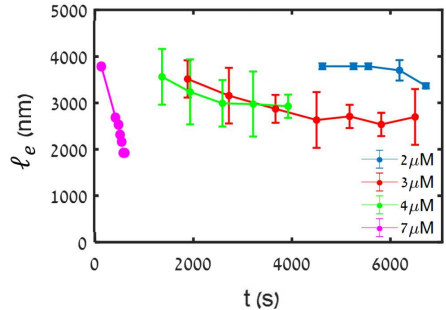}
		\caption{The entanglement length ($\ell_e$) as a function of time for samples with low actin monomer concentration. $\ell_e$ decreases by increasing actin monomer concentration.}
		\label{fig:Le_te_err_all}
	\end{figure}

	By knowing the mechanical properties of the forming network at each time point, the entanglement length $l_e$ can be estimated according to \cite{ent1,ent2}:
	\begin{equation}
	{\tau_e=\frac{1}{\omega_e}\cong\frac{\eta_0 \ell_e^4}{k_B Tl_p}},
	\label{Eq:le}
	\end{equation}
	where $\tau_e$ and $\omega_e$ are the entanglement time and frequency respectively. We note that we define  $\omega_e$ as the crossover frequency in which $G''$ becomes larger than $G'$ (Fig.~\ref{fig:All_freq_range}e). This choice is slightly different than the actual definition of  $\omega_e$, but is easier to extract in a consistent manner from our experiments. We extracted $\omega_e$ only for the samples in which this crossover was observed, namely for low initial actin monomer concentrations. The evolution of $\ell_e$ with time for these concentrations is presented in Fig.~\ref{fig:Le_te_err_all}.
	As expected the entanglement length, $\ell_e$, decreases with time as the filaments grow and the mesh size decreases. Furthermore, we find that the entanglement length decreases upon increasing monomer concentration, as the filament density increases and the mesh size decreases \cite{SCHMIDT1989}.
	
	\section{Conclusions}
	In this paper we have presented a new, indirect way of characterizing the kinetics of entangled F-actin networks formation. By using time resolved 1P microrheology, we obtained the evolution of the mechanical properties of a forming network with time resolution of 30s, for initial actin monomer concentrations varying from 2 to 24 $\mu M$.
	We found that the kinetics of the self-organization process is governed by nucleation at low actin monomer concentrations, with an indication that the stable nucleus is made of three monomers. This result is in accord with previous reports \cite{peryne1983}. At higher monomer concentrations, we found that the kinetics of self-organization is limited by diffusion corresponding to a pseudo-first order process suggested in earlier reports \cite{POLLARD2003,Hikoichi1983}. We assume that in these higher actin monomer concentrations nucleation occurs rapidly. 
	
	Our characterization technique enabled a unique measurement of the self organization kinetics past the formation of filaments, providing insight into how networks reorganize into an equilibrium structure once the initial polymerization is finished. Interestingly, we find that the shear modulus overshoots its final value at concentrations where polymerization is rapid. More striking, is the universal power law of -2/5 we find for the relaxation of the storage modulus to the equilibrium value. We assume the relaxation is related to actin filaments rearrangements, however, complementary experiments using cryo-electron microscopy might shed light on this relaxation process.
	
	These experiments set up the stage for studying the polymerization kinetics of actin networks of different structures, such as the branched network structure obtained in the presence of the arp2/3 protein. Experiments studying the effect of the presence of myosin motors on actin polymerization kinetics using the same methodology are presently underway. Our simplified in-vitro approach furthers the understanding of how underling physical principles contribute to the cytoskeleton active self-organization process.

	\section*{Conflicts of interest}
	There are no conflicts to declare.
	
	\section*{Supplementary Material}
	See \dag{ESI} material for ATP regeneration control experiment (Fig. S1), 2P microrheology complementary analysis (Fig. S2,S3), complementary data to Fig.~\ref{fig:All_freq_range} (Fig. S4) and theoretical values for steady state actin network in all actin monomer concentrations and theoretical values related to Fig.~\ref{fig:All_freq_range} (Tables S1 and S2 respectively).         
	
	\section*{Acknowledgements}
	We acknowledge financial support from the United States - Israel Binational Science Foundation (BSF) grant 2014314. Raya Sorkin acknowledges support through HFSP postdoctoral fellowship LT000419/2015.
	\providecommand*{\mcitethebibliography}{\thebibliography}
	\csname @ifundefined\endcsname{endmcitethebibliography}
	{\let\endmcitethebibliography\endthebibliography}{}

\end{document}